\documentclass[12pt,a4paper]{article}
\usepackage[latin1]{inputenc}
\usepackage{a4wide}
\usepackage{amsfonts}
\usepackage{amssymb}
\usepackage{amsmath}
\usepackage{bbm}
\usepackage{ifpdf}
\ifpdf
\usepackage[pdftex]{graphicx}
\usepackage[pdftex,unicode,implicit]{hyperref}
\hypersetup{%
  pdftitle    = {The supersymmetric solutions and extensions of
ungauged matter-coupled N=1,d=4 supergravity},
  pdfkeywords = {Supersymmetry, supergravity, BPS solutions},
  pdfauthor   = {Tomás Ortín},
  pdfcreator  = {pdf\LaTeXe\ with package \flqq hyperref\frqq},
  pdfproducer = {pdf\LaTeXe\ with package \flqq hyperref\frqq},
  pdfpagemode = None,  
  pdffitwindow= true,  
  unicode     = true,
  plainpages  = true,
  colorlinks  = true,  
  citecolor   = blue,
  urlcolor    = red,
  linkcolor   = black
}
\newcommand{\hepth}[1]{arXiv:{\tt
\href{http://www.arXiv.org/abs/hep-th/#1}{hep-th/#1}}}

\newcommand{\arxiv}[1]{{\tt
\href{http://www.arXiv.org/abs/#1}{arXiv:#1}}}
\else
  \usepackage[dvips]{graphicx}
  \usepackage[unicode,implicit]{hyperref}
  \newcommand{\hepth}[1]{arXiv:{\tt hep-th/#1}}

  \newcommand{\arxiv}[1]{{\tt arXiv:#1}}
\fi
\makeatletter
\@addtoreset{equation}{section}
\makeatother

\makeindex
\begin{document}

\begin{flushright}
\small
IFT-UAM/CSIC-08-08\\
\texttt{arXiv:0802.1799}\\
February $13^{\rm th}$ 2008\\
\normalsize
\end{flushright}

\begin{center}

\vspace{.7cm}

{\LARGE {\bf The supersymmetric solutions and\\[.5cm] extensions of
ungauged matter-coupled\\[.5cm] $N=1,d=4$ supergravity}}

\vspace{2cm}

\begin{center}
{\large\bf Tom\'{a}s Ort\'{\i}n}\\[.3cm]
\texttt{Tomas.Ortin@uam.es}\\[.8cm]
\textit{Instituto de F\'{\i}sica Te\'{o}rica UAM/CSIC, \\ Facultad de
Ciencias C-XVI, C.U.~Cantoblanco, E-28049 Madrid, Spain}
\end{center}

\vspace{4cm}

{\bf Abstract}

\begin{quotation}

  We find the most general supersymmetric solutions of ungauged $N=1,d=4$
  supergravity coupled to an arbitrary number of vector and chiral
  supermultiplets, which turn out to be essentially $pp$-waves and strings. We
  also introduce magnetic 1-forms and their supersymmetry transformations and
  2-forms associated to the isometries of the scalar manifold and their
  supersymmetry transformations. Only the latter can couple to BPS objects
  (strings), in agreement with our results.

\end{quotation}

\end{center}

\newpage
\pagestyle{plain}


\newpage

\section{Introduction}

Supersymmetric classical solutions of supergravity theories (low-energy
superstring theories) are a key tool in the current research on many topics
ranging from $AdS/CFT$ correspondence to stringy black-hole physics. Not all
locally supersymmetric solutions are necessarily interesting or useful in the
end, but, clearly, it is an important goal to find them all for every possible
supergravity theory.

This goal has been pursued and reached in several lower-dimensional theories
and families of theories. The pioneering work \cite{Tod:1983pm} was done in
1983 by Tod in pure, ungauged, $N=2,d=4$ supergravity. It has been
subsequently extended to the gauged case in Ref.~\cite{Caldarelli:2003pb}, to
include the coupling to general (ungauged) vector multiplets and
hypermultiplets in Refs.~\cite{Meessen:2006tu} and \cite{Huebscher:2006mr},
respectively and some partial results on the theory with gauged vector
multiplets have been recently obtained \cite{Huebscher:2007hj}.  Research on
pure $N=4,d=4$ supergravity was started in Ref.~\cite{Tod:1995jf} and
completed in Ref.~\cite{Bellorin:2005zc}.

In $d=5$, the minimal $N=1$ (sometimes referred as $N=2$) theory was worked
out in Ref.~\cite{Gauntlett:2002nw} and the results were extended to the
gauged case in Ref.~\cite{Gauntlett:2003fk}. The coupling to an arbitrary
number of vector multiplets and their Abelian gaugings was considered in
Refs.~\cite{Gutowski:2004yv,Gutowski:2005id}\footnote{Previous work on these
  theories can be found in Refs.~\cite{Chamseddine:1998yv,Sabra:1997yd}.}.
The inclusion of (ungauged) hypermultiplets was considered in
\cite{Bellorin:2006yr}\footnote{Previous partial results on that problem were
  presented in Refs.~\cite{Cacciatori:2002qx,Celi:2003qk,Cacciatori:2004qm}.}
and the extension to the most general gaugings with vector multiplets and
hypermultiplets was worked out in \cite{Bellorin:2007yp}.

The minimal $d=6$ SUGRA was dealt with in
Refs.~\cite{Gutowski:2003rg,Chamseddine:2003yy}, some gaugings were considered
in Ref.~\cite{Cariglia:2004kk} and the coupling to hypermultiplets has been
fully solved in Ref.~\cite{Jong:2006za}. 

All these works are essentially based on the method pioneered by Tod and
generalized by Gauntlett \textit{et al.} in Ref.~\cite{Gauntlett:2002nw},
which we will use here. An alternative method is that of spinorial geometry,
developed in Ref.~\cite{Gillard:2004xq}. Further works on this subject
  in 4 or higher dimensions are Refs.~\cite{Gauntlett:2002fz}.

It is somewhat surprising that the simpler $N=1,d=4$ theories have not yet
been studied. The purpose of this paper is to start filling this gap. We will
find all the supersymmetric configurations and solutions of ungauged $N=1,d=4$
supergravity and we will relate them to supersymmetric solutions of $N=2,d=4$
supergravity theories that we can truncate to $N=1,d=4$ theories following
\cite{Andrianopoli:2001zh,Andrianopoli:2001gm}. As we are going to see, there
are no timelike supersymmetric solutions such as charged, extreme, black holes
in these theories and in the null class we find essentially $pp$-waves, cosmic
strings and combinations of both. This is, precisely, the kind of
supersymmetric solutions of $N=2,d=4$ supergravity that would survive the
truncation to $N=1$.

We are also going to study the extension of the set of standard bosonic fields
of $N=1,d=4$ supergravity along the lines of Ref.~\cite{Bergshoeff:2007ij}. We
are going to show that we can add consistently (we can define supersymmetry
transformations for them such that the local supersymmetry algebra closes) the
magnetic vectors and also 2-forms which are associated to the isometries of
the scalar manifold. The electric and magnetic vectors of the theory transform
into the gauginos and not into the gravitino. This makes it impossible to
write a $\kappa$-symmetric action for 0-branes, in agreement with the absence
of supersymmetric black-hole solutions in the theory. The 2-forms do transform
into the gravitino and one can, in principle, construct $\kappa$-symmetric
actions for 1-branes, which agrees with the existence of supersymmetric string
solutions.

This paper is organized as follows: in Section~\ref{sec-N1d4sugra} we
introduce ungauged $N=1,d=4$ supergravity coupled to vector and chiral
supermultiplets. We obtain this theory by truncation of ungauged $N=2,d=4$
supergravity coupled to vector supermultiplets and hypermultiplets in
Appendix~\ref{app-truncation}. This helps us to fix the conventions and to
relate the solutions to $N=2,d=4$ solutions. In Section~\ref{sec-setup} we set
up the problem we aim to solve. In Section~\ref{sec-configurations} we find
all the bosonic field configurations that admit Killing spinors (as we check
in Section~\ref{sec-KSEs}) and in Section~\ref{sec-solutions} we identify
amongst them those that satisfy the classical equations of motion, which
solves our problem. In Section~\ref{sec-extensions} we find the bosonic field
extensions of the theory. Finally, in Section~\ref{sec-conclusions} we discuss
our results and give our conclusions.

After completion of this work we became aware that a similar results have been
obtained by U.~Gran, J.~Gutowski and G.~Papadopoulos and are about to be
published \cite{Gran:2008vx}.


\section{Matter-coupled, ungauged,  $N=1, d=4$ supergravity}
\label{sec-N1d4sugra}

In this section we describe briefly the theory \cite{Cremmer:1982en}, which is
obtained by truncation of $N=2,d=4$ theories in the appendix.  Our conventions
are derived from those we use in the study of $N=2,d=4$ theories
\cite{Meessen:2006tu,Huebscher:2006mr,Huebscher:2007hj}. It contains a
supergravity multiplet with one graviton $e^{a}{}_{\mu}$ and one chiral
gravitino $\psi_{\bullet\mu}$, $n_{C}$ chiral multiplets with as many chiral
dilatini $\chi_{\bullet}{}^{i}$ and complex scalars $Z^{i}$, $i=1,\cdots
n_{C}$ that parametrize a K\"ahler-Hodge manifold with metric
$\mathcal{G}_{ij^{*}}$, and $n_{V}$ vector multiplets with as many vector
fields $A^{\Lambda}$ and chiral gaugini $\lambda_{\bullet}{}^{\Lambda}$
$\Lambda=1,\cdots,n_{V}$.

The action for the bosonic fields is 

\begin{equation}
\label{eq:actionN1}
 S  =  {\displaystyle\int} d^{4}x \sqrt{|g|}
\left[R +2\mathcal{G}_{ij^{*}}\partial_{\mu}Z^{i}
\partial^{\mu}Z^{*\, j^{*}} 
-\Im{\rm m}f_{\Lambda\Sigma} 
F^{\Lambda\, \mu\nu}F^{\Sigma}{}_{\mu\nu}
-\Re{\rm e}f_{\Lambda\Sigma} 
F^{\Lambda\, \mu\nu}{}^{\star}F^{\Sigma}{}_{\mu\nu}
\right]\, ,
\end{equation}

\noindent
where $f_{\Lambda\Sigma}(Z)$ is a $n_{V}\times n_{V}$ matrix with entries
which are holomorphic functions of the complex scalars and with definite
positive imaginary part. 

The supersymmetry transformation rules for the bosonic fields are

\begin{eqnarray}
\label{eq:susytranseN1}
\delta_{\epsilon} e^{a}{}_{\mu} & = &  
-{\textstyle\frac{i}{4}} \bar{\psi}_{\bullet\, \mu}\gamma^{a}\epsilon^{\bullet}
+\mathrm{c.c.}\, ,\\
& & \nonumber \\
\label{eq:susytransAN1}
\delta_{\epsilon} A^{\Lambda}{}_{\mu} & = & 
{\textstyle\frac{i}{8}}
\bar{\lambda}_{\bullet}{}^{\Lambda}\gamma_{\mu}
\epsilon^{\bullet}
+\mathrm{c.c.}\, ,\\
& & \nonumber \\
\delta_{\epsilon} Z^{i} & = & 
{\textstyle\frac{1}{4}} \bar{\chi}_{\bullet}{}^{i}\epsilon_{\bullet}\, ,
\label{eq:susytransZN1}
\end{eqnarray}

\noindent
and those of the fermions, for vanishing fermions, are

\begin{eqnarray}
\delta_{\epsilon}\psi_{\bullet\, \mu} & = & 
\mathfrak{D}_{\mu}\epsilon_{\bullet}
=
\left(\nabla_{\mu} +{\textstyle\frac{i}{2}} \mathcal{Q}_{\mu}\right)\epsilon_{\bullet} \, ,
\label{eq:gravisusyruleN1}\\
& & \nonumber \\
\delta_{\epsilon}\lambda_{\bullet}{}^{\Lambda} & = &
{\textstyle\frac{1}{2}}\not\!
F^{\mathbf{\Lambda}+}\epsilon_{\bullet}\, ,
\label{eq:gaugsusyruleN1} \\
& & \nonumber \\
\delta_{\epsilon}\chi_{\bullet}{}^{i} & = & 
i\not\!\partial Z^{i}\epsilon^{\bullet}\, ,
\label{eq:dilasusyruleN1}
\end{eqnarray}

\noindent
where $\mathcal{Q}_{\mu}$ is the pullback of the K\"ahler 1-form connection

\begin{equation}
\label{eq:K1form}
\mathcal{Q} \equiv {\textstyle\frac{1}{2i}}(dz^{i}\partial_{i}\mathcal{K} -
dz^{*\, i^{*}}\partial_{i^{*}}\mathcal{K})\, ,
\end{equation}

\noindent
where $\mathcal{K}$ is the K\"ahler potential from which the K\"ahler metric
can be derived in the standard fashion, namely

\begin{equation}
\mathcal{G}_{ij^{*}}=
 \partial_{i}\partial_{j^{*}}\mathcal{K}\, .
\end{equation}

For convenience, we denote the bosonic equations of motion by

\begin{equation}
\mathcal{E}_{a}{}^{\mu}\equiv 
-\frac{1}{2\sqrt{|g|}}\frac{\delta S}{\delta e^{a}{}_{\mu}}\, ,
\hspace{.5cm}
\mathcal{E}^{i} \equiv -\frac{\mathcal{G}^{ij^{*}}}{2\sqrt{|g|}}
\frac{\delta S}{\delta Z^{*j^{*}}}\, ,
\hspace{.5cm}
\mathcal{E}_{\Lambda}{}^{\mu}\equiv 
\frac{1}{4\sqrt{|g|}}\frac{\delta S}{\delta A^{\Lambda}{}_{\mu}}\, .
\end{equation}

\noindent
and the Bianchi identities for the vector field strengths by

\begin{equation}
\label{eq:Bianchiidentities}
\mathcal{B}^{\Lambda\, \mu} \equiv \nabla_{\nu}\star F^{\Lambda\,
  \nu\mu}\, ,
\,\,\,\,\, 
\star\mathcal{B}^{\Lambda}\equiv -dF^{\Lambda}\, .
\end{equation}

Then, using the action Eq.~(\ref{eq:actionN1}), we find

\begin{eqnarray}
\mathcal{E}_{\mu\nu} & = & 
G_{\mu\nu}
+2\mathcal{G}_{ij^{*}}[\partial_{\mu}Z^{i} \partial_{\nu}Z^{*\, j^{*}}
-{\textstyle\frac{1}{2}}g_{\mu\nu}
\partial_{\rho}Z^{i}\partial^{\rho}Z^{*\, j^{*}}]\nonumber \\
& & \nonumber \\
& & 
-4\Im {\rm m}f_{\Lambda\Sigma}
F^{\Lambda\, +}{}_{\mu}{}^{\rho}F^{\Sigma\, -}{}_{\nu\rho}\, ,
\label{eq:Emn}\\
& & \nonumber \\
\mathcal{E}_{i} & = & \mathcal{G}_{ij^{*}}\mathfrak{D}_{\mu}
\partial^{\mu}Z^{*\, i^{*}}
+\partial_{i}[
F_{\Lambda}{}^{\mu\nu}\star F^{\Lambda}{}_{\mu\nu}]
\\
& & \nonumber \\
& = & 
\mathcal{G}_{ij^{*}}\mathfrak{D}_{\mu}
\partial^{\mu}Z^{*\, i^{*}} 
-{\textstyle\frac{i}{2}}
\partial_{i}f_{\Lambda\Sigma}F^{\Lambda\, +}{}_{\mu\nu}
F^{\Sigma\, +\, \mu\nu}\, ,
\label{eq:Ei}\\
& & \nonumber \\
\mathcal{E}_{\Lambda}{}^{\mu} & = & 
\nabla_{\nu}\star F_{\Lambda}{}^{\nu\mu}\, ,
\label{eq:ERm}
\end{eqnarray}

\noindent
where we have defined the dual vector field strength $F_{\Lambda}$
by 

\begin{equation}
\label{eq:dualF}
F_{\Lambda\, \mu\nu} \equiv   
-\frac{1}{2\sqrt{|g|}}\frac{\delta S}{\delta {}^{\star}F^{\Lambda}{}_{\mu\nu}}
= \Re {\rm e}f_{\Lambda\Sigma}F^{\Sigma}{}_{\mu\nu}
-\Im {\rm m}f_{\Lambda\Sigma}{}^{*}F^{\Sigma}{}_{\mu\nu}
=2\Re{\rm e}\, (f_{\Lambda\Sigma}F^{\Sigma\, +})\, .
\end{equation}

The Maxwell equations can be read as Bianchi identities for these dual field
strengths ensuring the local existence of $n_{V}$ dual vector potentials
$A_{\Lambda}$ such that 

\begin{equation}
F_{\Lambda}=dA_{\Lambda}\, .  
\end{equation}

It is convenient to combine the standard, electric, field strengths and
potentials and their duals Eq.~(\ref{eq:dualF}) into a single
$2n_{V}$-dimensional symplectic vector

\begin{equation}
\mathcal{F}\equiv 
\left(
  \begin{array}{c}
   F^{\Lambda} \\ F_{\Lambda} \\ 
  \end{array}
\right)
= d \mathcal{A} \equiv 
d
\left(
  \begin{array}{c}
   A^{\Lambda} \\ A_{\Lambda} \\ 
  \end{array}
\right)\, .  
\end{equation}

The global symmetries of these theories will be the isometries of the scalar
manifold that can be embedded in $Sp(2n_{V},\mathbb{R})$
\cite{Gaillard:1981rj}.


\section{Supersymmetric configurations: general setup}
\label{sec-setup}

Our first goal is to find all the bosonic field configurations
$\{g_{\mu\nu},F^{\Lambda}{}_{\mu\nu}, Z^{i}\}$ for which the Killing
spinor equations (KSEs):

\begin{eqnarray}
\delta_{\epsilon}\psi_{\bullet\, \mu} & = & 
\mathfrak{D}_{\mu}\epsilon_{\bullet}
 =0 \, ,
\label{eq:KSE1}\\
& & \nonumber \\
\delta_{\epsilon}\lambda_{\bullet}{}^{\Lambda} & = &
{\textstyle\frac{1}{2}}\not\!
F^{\mathbf{\Lambda}+}\epsilon_{\bullet}=0\, ,
\label{eq:KSE2} \\
& & \nonumber \\
\delta_{\epsilon}\chi_{\bullet}{}^{i} & = & 
i\not\!\partial Z^{i}\epsilon^{\bullet}=0\, ,
\label{eq:KSE3}
\end{eqnarray}

\noindent
admit at least one solution. It must be stressed that the configurations
considered need not be classical solutions of the equations of motion.
Furthermore, we will not assume that the Bianchi identities are satisfied by
the field strengths of a configuration.

Our second goal will be to identify among all the supersymmetric field
configurations those that satisfy all the equations of motion (including
the Bianchi identities).

Let us initiate the analysis of the KSEs by studying their integrability
conditions.


\subsection{Killing Spinor Identities (KSIs)}

Using the supersymmetry transformation rules of the bosonic fields
Eqs.~(\ref{eq:susytranseN1}--\ref{eq:susytransZN1}) and
using the results of Refs.~\cite{Kallosh:1993wx,Bellorin:2005hy} we can derive
following relations (\textit{Killing spinor identities}, KSIs) between the
(off-shell) equations of motion of the bosonic fields
Eqs.~(\ref{eq:Emn}--\ref{eq:ERm}) that are satisfied by any
field configuration $\{e^{a}{}_{\mu},A^{\Lambda}{}_{\mu},Z^{i}\}$ admitting
Killing spinors:

\begin{eqnarray}
\mathcal{E}^{\mu}{}_{a}\gamma^{a}\epsilon^{\bullet} & = & 0\, ,\\
& & \nonumber \\
\mathcal{E}_{\Lambda}{}^{\mu}\gamma_{\mu}\epsilon^{\bullet} & = & 0\, ,\\
& & \nonumber \\
\mathcal{E}_{i}\epsilon_{\bullet} & = & 0\, . 
\end{eqnarray}

In this way of finding the KSIs the Bianchi identities are assumed to be
satisfied. It is convenient to have KSIs in which they appear explicitly.
These can be found through the integrability conditions of the KSEs.  The only
KSI in which we expect the Bianchi identities to appear is the second one
above, which involves the Maxwell equations. The Bianchi identities should
combine with the Maxwell equations in a electric-magnetic duality-invariant
way. Then, the second KSI above should be replaced by 

\begin{equation}
(\mathcal{E}_{\Lambda}{}^{\mu}-f_{\Lambda\Sigma}\mathcal{B}^{\Sigma\, \mu})
 \gamma_{\mu}\epsilon^{\bullet} = 0\, .  
\end{equation}

This can be explicitly checked via the following integrability condition of
the gaugini:

\begin{equation}
  \begin{array}{rcl}
\not\!\!\mathfrak{D}\delta_{\epsilon}\lambda_{\bullet}{}^{\Lambda} & = & 
(\Im{\rm m} f)^{-1|\Lambda\Sigma}
(\not\!\mathcal{E}_{\Sigma}-f^{*}_{\Sigma\Omega}\not\!\mathcal{B}^{\Omega})
\epsilon_{\bullet}\\
& & \\
& & 
+i(\Im{\rm m} f)^{-1|\Lambda\Sigma}
\not\!\partial f_{\Sigma\Omega}\delta_{\epsilon}\lambda_{\bullet}{}^{\Omega}
-{\textstyle\frac{1}{4}}\not\! F^{\Lambda\, -}
\delta_{\epsilon}\chi^{\bullet\, i^{*}}
+{\textstyle\frac{1}{2}}\gamma^{\mu}\not\! F^{\Lambda\, +}
\delta_{\epsilon}\psi_{\bullet\, \mu}\, .\\
\end{array}
\end{equation}

From these identities one can derive identities that involve tensors
constructed as bilinears of the Killing spinors.  In $N=1$ supergravity there
is only one chiral spinor $\epsilon_{\bullet}$. With it, we can only construct
a real null vector $l_{\mu} =
i\sqrt{2}\bar{\epsilon}^{\bullet}\gamma_{\mu}\epsilon_{\bullet}$, one
self-dual 2-form $\Phi_{\mu\nu}=
\bar{\epsilon}_{\bullet}\gamma_{\mu\nu}\epsilon_{\bullet}$ and no scalars. In
the $N>1$ cases one can construct a vector which is non-spacelike and, thus,
one considers separately the case in which the vector is timelike and the case
in which it is null. In $N=1,d=4$ there is no timelike case.  It is convenient
to introduce an auxiliary chiral spinor $\eta_{\bullet}$ with normalization

\begin{equation}
\label{eq:etanorm}
\bar{\epsilon}_{\bullet}\eta_{\bullet}={\textstyle\frac{1}{2}}\, ,
\end{equation}

\noindent
and with the same chirality but opposite K\"ahler weight as
$\epsilon_{\bullet}$. With both spinors we construct the null tetrad

\begin{equation}
\label{eq:nulltetraddef}
\begin{array}{rclrcl}
l_{\mu} & = & i\sqrt{2}\bar{\epsilon}^{\bullet}\gamma_{\mu}\epsilon_{\bullet}\, ,
\hspace{1.5cm} &
n_{\mu} & = & i\sqrt{2}\bar{\eta}^{\bullet}\gamma_{\mu}\eta_{\bullet}\, ,\\
& & & & & \\
m_{\mu} & = & i\sqrt{2}\bar{\epsilon}^{\bullet}\gamma_{\mu}\eta_{\bullet}\, , 
\hspace{1.5cm} &
m_{\mu}^{*} & = & i\sqrt{2}\bar{\epsilon}_{\bullet}\gamma_{\mu}\eta^{\bullet}\, .\\
\end{array}
\end{equation}

\noindent
$l$ and $n$ have 0 $U(1)$ charges but $m$ has $-2$ times the charges of
$\epsilon$ and $m^{*}$ has $+2$ times the charges of $\epsilon$.

\begin{eqnarray}
\mathcal{E}_{\mu\nu}l^{\nu}= \mathcal{E}_{\mu\nu}m^{\nu} & = & 0\, ,\\
& & \nonumber \\
(\mathcal{E}_{\Lambda\, \mu}
-f_{\Lambda\Sigma}\mathcal{B}^{\Sigma}{}_{\mu})l^{\mu} =
(\mathcal{E}_{\Lambda\, \mu}
-f_{\Lambda\Sigma}\mathcal{B}^{\Sigma}{}_{\mu})
  m^{\mu} 
& = & 0\, ,\\
& & \nonumber \\
\mathcal{E}_{i} & = & 0\, .  
\end{eqnarray}

\noindent
This means that the only independent equations of motion that we have to
impose on supersymmetric configurations are 

\begin{eqnarray}
\label{eq:eom1}
\mathcal{E}_{\mu\nu}n^{\mu}n^{\nu} & = & 0 \, ,\\
& & \nonumber \\
\label{eq:eom2}
(\mathcal{E}_{\Lambda\, \mu}
-f_{\Lambda\Sigma}\mathcal{B}^{\Sigma}{}_{\mu})n^{\mu} & = & 0\, ,\\
& & \nonumber \\
\label{eq:eom3}
(\mathcal{E}_{\Lambda\, \mu}
-f_{\Lambda\Sigma}\mathcal{B}^{\Sigma}{}_{\mu})m^{*\mu} & = & 0\, .
\end{eqnarray}


\section{Supersymmetric configurations and solutions}
\label{sec-configurationsandsolutions}


\subsection{Supersymmetric configurations}
\label{sec-configurations}

Our first goal is to derive from the KSEs consistency conditions expressed in
terms of the null tetrad vectors.

Acting on the KSE Eq.~(\ref{eq:KSE2}) with
$\bar{\epsilon}^{\bullet}\gamma_{\mu}$ and $\bar{\eta}^{\bullet}\gamma_{\mu}$
we get, respectively

\begin{eqnarray}
F^{\Lambda\, +}{}_{\mu\nu}l^{\nu} & = & 0\, ,\\
& & \nonumber \\
F^{\Lambda\, +}{}_{\mu\nu}m^{*\nu} & = & 0\, ,
\end{eqnarray}

\noindent
which imply that 

\begin{equation}
\label{eq:F+}
F^{\Lambda\, +} = 
{\textstyle\frac{1}{2}}\phi^{\Lambda}\hat{l} \wedge \hat{m}^{*}\, ,  
\end{equation}

\noindent
for some functions $\phi^{\Lambda}$ to be determined.  This form of
$F^{\Lambda\, +}$ solves the KSE Eq.~(\ref{eq:KSE2}) by virtue of the Fierz
identities

\begin{equation}
l_{\mu}\gamma^{\mu\nu}\epsilon_{\bullet}=l^{\nu}\epsilon_{\bullet}\, ,
\hspace{1cm}
m^{*}_{\mu}\gamma^{\mu\nu}\epsilon_{\bullet}=m^{*\nu}\epsilon_{\bullet}\, .
\end{equation}

Acting now on the KSE Eq.~(\ref{eq:KSE3}) with $\bar{\epsilon}_{\bullet}$ and
$\bar{\eta}_{\bullet}$ we get, respectively

\begin{eqnarray}
\label{eq:ldZ}
l^{\mu}\partial_{\mu}Z^{i} & = & 0\, ,\\
& & \nonumber \\
\label{eq:mdZ}  
m^{\mu}\partial_{\mu}Z^{i} & = & 0\, ,
\end{eqnarray}

\noindent
which imply 

\begin{equation}
\label{eq:dZ}
dZ^{i} = A^{i}\hat{l} +B^{i}\hat{m}\, ,
\end{equation}

\noindent
for some functions $A^{i}$ and $B^{i}$ to be determined. This form of $dZ^{i}$
solves  the KSE Eq.~(\ref{eq:KSE3}) by virtue of the Fierz identities

\begin{equation}
\label{eq:mlFierz}
\not l\epsilon^{*}=\not\!m\epsilon^{*}=0\, .  
\end{equation}

Now, , from the normalization condition of the auxiliary spinor
$\eta_{\bullet}$ we find the condition

\begin{equation}
\label{eq:deta}
  \mathfrak{D}_{\mu}\eta_{\bullet} + a_{\mu}\epsilon_{\bullet} =0\, ,
\end{equation}

\noindent
for some $a_{\mu}$ with $U(1)$ charges $-2$ times those of $\epsilon$, {\em
  i.e.}

\begin{equation}
\mathfrak{D}_{\mu}a_{\nu} =
(\nabla_{\mu} -i\mathcal{Q}_{\mu})a_{\nu}\, ,
\end{equation}

\noindent
to be determined by the requirement that the integrability conditions
of this differential equation have to be compatible with those of the
differential equation for $\epsilon$.

Taking the covariant derivative of the null tetrad vectors and using 
the KSE Eq.~(\ref{eq:KSE1}), we find

\begin{eqnarray}
\label{eq:dtetrad1}
\mathfrak{D}_{\mu} l_{\nu} & = &\nabla_{\mu} l_{\nu}  =0\, ,\\
& & \nonumber \\
\label{eq:dtetrad2}
\mathfrak{D}_{\mu} n_{\nu} & = & \nabla_{\mu}n_{\nu}=
-a^{*}_{\mu}m_{\nu} -a_{\mu}m^{*}_{\nu}\, ,\\
& & \nonumber \\
\label{eq:dtetrad3}
\mathfrak{D}_{\mu} m_{\nu} & = & 
(\nabla_{\mu} -i\mathcal{Q}_{\mu})m_{\nu} 
=-a_{\mu}l_{\nu}\, .
\end{eqnarray}

\noindent
The first of these equations is solved by identifying the most general metric
compatible with it: a Brinkmann $pp$-wave metric \cite{kn:Br1,kn:Br2}. One
introduces the coordinates $u$ and $v$ such that

\begin{eqnarray}
\hat{l}= l_{\mu}dx^{\mu} & \equiv & du\, , \label{u}\\
& & \nonumber \\
l^{\mu}\partial_{\mu}  & \equiv & \frac{\partial}{\partial v}\, , \label{v}
\end{eqnarray}

\noindent
and defines a complex coordinate $z$ by 

\begin{equation}
\label{eq:z}
\hat{m} = e^{U}dz\, ,  
\end{equation}

\noindent
where $U$ may depend on $z,z^{*}$ and $u$. The most general form that $\hat{n}$
can take in this case is

\begin{equation}
\hat{n}= dv + H du +\hat{\omega}\, ,   
\hspace{1cm}
\hat{\omega}=\omega_{\underline{z}}dz +\omega_{\underline{z}^{*}}dz^{*}\, ,
\end{equation}

\noindent
where all the functions in the metric are independent of $v$ and where
either $H$ or the 1-form $\hat{\omega}$ could, in principle, be removed by a
coordinate transformation but we have to check that the tetrad
integrability equations (\ref{eq:dtetrad1})-(\ref{eq:dtetrad3}) are
satisfied by our choices of $e^{U},H$ and $\hat{\omega}$

The above choice of coordinates leads to the metric

\begin{equation}
\label{eq:Brinkmetric}
ds^{2} = 2 du (dv + H du +\hat{\omega})
-2e^{2U}dzdz^{*}\, .
\end{equation}

\noindent
It also implies that the complex scalars $Z^{i}$ are functions of $z$ and $u$
but not of $z^{*}$ and $v$. The same is true for $A^{i}$ and $B^{i}$.

Let us consider the tetrad integrability equations
(\ref{eq:dtetrad1})-(\ref{eq:dtetrad3}): the first equation is solved because
the metric does not depend on $v$. The third equation, with the choice of
coordinate $z$, Eq.~(\ref{eq:z}), implies

\begin{eqnarray}
\hat{a} & = & n^{\mu}
(\partial_{\mu}U -i\mathcal{Q}_{\mu})\hat{m}+D\hat{l}\, ,\\
& & \nonumber \\
\label{eq:mDU}
m^{\mu}\partial_{\mu}(U-i \mathcal{Q}_{\mu}) & = & 0\, ,
\end{eqnarray}

\noindent
where $D$ is a function to be determined.

The second equation can be written using the definition of the K\"ahler
connection and the dependence $Z^{i}(z,u)$ in the form

\begin{equation}
\label{eq:mDU2}
\partial_{\underline{z}^{*}}(U+\mathcal{K}/2) = 0\,\,\, \Rightarrow
U=-\mathcal{K}/2+h(u)\, , 
\end{equation}

\noindent
where $h(u)$ can be eliminated by a coordinate redefinition that does not
change the general form of the Brinkmann metric.

The second tetrad integrability equation (\ref{eq:dtetrad2}) implies

\begin{eqnarray}
D & = & e^{-U}(\partial_{\underline{z}^{*}}H 
-\dot{\omega}_{\underline{z}^{*}})\, ,\\
& & \nonumber \\
\label{eq:doQ}
(d\omega)_{\underline{z}\underline{z}^{*}} & = & 
2ie^{2U}n^{\mu}\mathcal{Q}_{\mu}\, ,
\end{eqnarray}

\noindent
whence $\hat{a}$ is given by

\begin{equation}
\label{eq:exprehata}
\hat{a} =   [\dot{U} -{\textstyle\frac{1}{2}}e^{-2U}
(d\omega)_{\underline{z}\underline{z}^{*}}]\hat{m}
+e^{-U}(\partial_{\underline{z}^{*}}H 
-\dot{\omega}_{\underline{z}^{*}})\hat{l}\, .
\end{equation}


\subsection{Killing spinor equations}
\label{sec-KSEs}

We are now going to see that field configurations given by a metric of the
form (Eqs.~(\ref{eq:Brinkmetric}) where $\hat{\omega}$ satisfies
(Eq.~(\ref{eq:doQ})) and $U$ satisfies Eq.~(\ref{eq:mDU2}), field strengths
given by Eqs.~(\ref{eq:F+}) and scalars of the form (\ref{eq:dZ}) are always
supersymmetric, even though we derived these equations as necessary conditions
for supersymmetry.

With the above form of the scalars and vector field strengths the KSE
$\delta_{\epsilon}\chi_{\bullet}{}^{i}=0$ takes the form

\begin{equation}
i[A^{i}\not l+B^{i}\not\!\! m ]\epsilon^{\bullet}=0\, .   
\end{equation}

\noindent
This equation is solved by imposing two conditions on the spinors:

\begin{equation}
\label{eq:nullconditions}
\not l \epsilon^{\bullet} = 0\, ,
\hspace{1cm}
\not\!\! m \epsilon^{\bullet} = 0\, .
\end{equation}

\noindent
As shown in Ref.~(\cite{Meessen:2006tu}) these two constraints are not just
compatible but equivalent and only half of the supersymmetries are broken by
them.

Let us now consider the KSE $\delta_{\epsilon}\psi_{\bullet\, a}=0$.  It takes
the form

\begin{equation}
\{\partial_{a} -{\textstyle\frac{1}{4}}\omega_{abc}\gamma^{bc}
+{\textstyle\frac{i}{2}}
\mathcal Q_{a}\}\epsilon_{\bullet}=0\, .
\end{equation}

The $v$ component is automatically satisfied for $v$-independent Killing
spinors. The $z$ and $z^{*}$ components take, after use of the constraints
Eq.~(\ref{eq:nullconditions}) and their consequence
$\gamma^{zz^{*}}\epsilon_{\bullet}=\epsilon_{\bullet}$ the form

\begin{eqnarray}
\{\partial_{\underline{z}} 
+{\textstyle\frac{1}{2}}\partial_{\underline{z}}(U+\mathcal{K}/2)\}
\epsilon_{\bullet} & = &
0\, ,\\
& & \nonumber \\
\{\partial_{\underline{z}^{*}} 
+{\textstyle\frac{1}{2}}\partial_{\underline{z}^{*}}(U+\mathcal{K}/2)\}
\epsilon_{\bullet} & = &
0\, .
\end{eqnarray}

\noindent
They are solved for $z$- and $z^{*}$-independent spinors once
Eq.~(\ref{eq:mDU2}) is taken into account. The $u$ component simply implies
that the Killing spinors are also $u$-independent. 

Thus, all the configurations identified are supersymmetric with Killing
spinors which are constant spinors satisfying Eqs.~(\ref{eq:nullconditions}).
Thus, they generically preserve $1/2$ of the supersymmetries (no less).


\subsection{Solutions}
\label{sec-solutions}

The Bianchi identities take, in differential-form language, the form

\begin{equation}
\hat{\mathcal{B}}^{\Sigma} =  -dF^{\Lambda} =
{\textstyle\frac{1}{2}}d(\phi^{\Sigma}\hat{m}+\mathrm{c.c})\wedge \hat{l}\, ,
\end{equation}

\noindent
and are solved by 

\begin{equation}
A^{\Lambda}= \varphi^{\Lambda}(z,u)du +\mathrm{c.c.}\, ,
\hspace{1cm}
e^{\mathcal{K}/2}\partial_{\underline{z}}\varphi^{\Lambda}(z,u) =
\phi^{*\Lambda}\, .
\end{equation}

The Maxwell equations take  the form

\begin{equation}
\hat{\mathcal{E}}_{\Lambda}= 
d (f_{\Lambda\Sigma}F^{\Lambda\, +}+ \mathrm{c.c.})=
-{\textstyle\frac{1}{2}}
d(f_{\Lambda\Sigma}\phi^{\Sigma}\hat{m}^{*}+\mathrm{c.c})\wedge \hat{l}\, ,
\end{equation}

\noindent
which is solved by holomorphic functions $\varphi_{\Lambda}(z,u)$ such that

\begin{equation}
\partial_{\underline{z}}\varphi_{\Lambda}(z,u) =
f^{*}_{\Lambda\Sigma}\phi^{*\Sigma}e^{-\mathcal{K}/2}\, .  
\end{equation}

\noindent
Using the solution of the Bianchi identities, we get

\begin{equation}
\partial_{\underline{z}}\varphi_{\Lambda}(z,u) =
f^{*}_{\Lambda\Sigma}  \partial_{\underline{z}}\varphi^{\Sigma}(z,u)\, .
\end{equation}

\noindent
Taking now into account that $f_{\Lambda\Sigma}$ is a holomorphic function of
the $Z^{i}$s which are, themselves, holomorphic functions of $z$ (and standard
functions of $u$), we arrive to the conclusion that the above equation can be
solved in two ways: either the $Z^{i}$s are $z$-independent or 

\begin{equation}
\label{eq:fphiconstraint}
\partial_{\underline{z}}\varphi_{\Lambda}(z,u) =
f^{*}_{\Lambda\Sigma}  \partial_{\underline{z}}\varphi^{\Sigma}(z,u)=0\, .
\end{equation}

\noindent
In general $f_{\Lambda\Sigma}$ will not have null eigenvectors and, therefore,
the only generic solutions are $z$-independent $\varphi^{\Sigma}$ and,
therefore, trivial vector fields.

Taking into account Eq.~(\ref{eq:fphiconstraint}), the only non-automatically
satisfied component of the Einstein equations is\footnote{For simplicity we
  choose the gauge $\omega=0$.}

\begin{equation}
\label{eq:EEuu}
\partial_{\underline{z}}\partial_{\underline{z}^{*}}H
-e^{-\mathcal{K}/2}\partial_{\underline{u}}^{2}e^{-\mathcal{K}/2} 
-e^{-\mathcal{K}}\mathcal{G}_{ij^{*}}
\partial_{\underline{u}}Z^{i}\partial_{\underline{u}}Z^{*\, j^{*}}
-{\textstyle\frac{1}{2}}\Im{\rm m}f_{\Lambda\Sigma} 
\partial_{\underline{z}}\varphi^{\Lambda}\partial_{\underline{z}^{*}}\varphi^{*\Sigma} 
=0\, .
\end{equation}

There are two cases to be considered:

\begin{itemize}

\item When the $Z^{i}$s are $z$-independent. Then

\begin{equation}
 H = \Re{\rm e}f(z)+
[  e^{-\mathcal{K}/2}\partial_{\underline{u}}^{2}e^{-\mathcal{K}/2} 
+e^{-\mathcal{K}}\mathcal{G}_{ij^{*}}
\partial_{\underline{u}}Z^{i}\partial_{\underline{u}}Z^{*\, j^{*}}]|z|^{2}
+{\textstyle\frac{1}{2}}\Im{\rm m}f_{\Lambda\Sigma} 
\varphi^{\Lambda}\varphi^{*\Sigma}\, . 
\end{equation}

These solutions describe gravitational, electromagnetic and scalar $pp$ waves.

\item When the $Z^{i}$s are not $z$-independent. The vector fields are
  trivial, but the above equation is not easy to integrate. In the special
  case in which the $Z^{i}$s are $u$-independent holomorphic functions of $z$

\begin{equation}
   H = \Re{\rm e}f(z)\, .  
\end{equation}

\end{itemize}

These solutions describe a superposition of a $pp$-wave and cosmic strings
such as those studied in
Refs.~\cite{Greene:1989ya,deAzcarraga:1989gm,Bergshoeff:2006jj,Bergshoeff:2007ij} and
found in $N=4,d=4$ \cite{Tod:1995jf,Bellorin:2005hy} and $N=2,d=4$
\cite{Meessen:2006tu,Huebscher:2006mr} theories.


\section{Extensions}
\label{sec-extensions}

In this section we are going to explore the possible extensions of the
standard formulation of $N=1,d=4$ supergravity, using our previous results on
the supersymmetric solutions of the theory. These suggest the possible
addition of 2-forms associated to the isometries of the K\"ahler scalar
manifold. These should couple to the cosmic string solutions exactly in the
form discussed in Ref.~\cite{Bergshoeff:2007ij} for $N=2,d=4$ supergravity. Since one
can define magnetic potentials from the Maxwell equations, it should also be
possible to add dual, magnetic, 1-forms. These, however, may not couple to any
standard 0-brane since all 1-forms transform into gaugini (and not the
gravitino) under supersymmetry.


\subsection{1-forms}
\label{sec-1-forms}

Given the supersymmetry transformation rule of the standard (electric)
potentials Eq.~(\ref{eq:susytransAN1}) and the definition of the dual field
strengths Eq.~(\ref{eq:dualF}), the simplest Ansatz for the transformation of
the dual (magnetic) potentials $A_{\Lambda}$ would be 

\begin{equation}
\label{eq:susytransAN1dualansatz}
\delta_{\epsilon} A_{\Lambda\, \mu} =
{\textstyle\frac{i}{8}}f^{*}_{\Lambda\Sigma}
\bar{\epsilon}_{\bullet}\gamma_{\mu}
\lambda^{\bullet\, \Sigma}
+\mathrm{c.c.}\, .
\end{equation}



\begin{equation}
  \begin{array}{rcl}
[\delta_{\eta},\delta_{\epsilon}] A_{\Lambda\, \mu} & = & 
-2\Re{\rm e}[a f^{*}_{\Lambda\Sigma}F^{\mathbf{\Sigma}-}{}_{\mu\nu}]
\xi^{\nu}\, , \\
\end{array}
\end{equation}

\noindent
where

\begin{equation}
\label{eq:xi}
\xi^{\nu}\equiv 
{\textstyle\frac{i}{4}}\bar{\epsilon}_{\bullet}\gamma^{\nu}\eta^{\bullet}
+\mathrm{c.c.}\, .  
\end{equation}

\noindent
In absence of the functions $f_{\Lambda\Sigma}$, we  have 

\begin{equation}
[\delta_{\eta},\delta_{\epsilon}] A^{\Lambda}{}_{\mu} = 
 -2\Re{\rm e}[F^{\Lambda\, -}{}_{\mu\nu}]\xi^{\nu}=
-F^{\Lambda}{}_{\mu\nu}\xi^{\nu} = [\delta_{\rm g.c.t.}(\xi)+\delta_{\rm
  gauge}(\Lambda^{\Lambda})]A^{\Lambda}{}_{\mu}\, ,
\end{equation}

\noindent
where 

\begin{equation}
\delta_{\rm g.c.t.}(\xi) A^{\Lambda}{}_{\mu} =
\xi^{\nu}\partial_{\nu}A^{\Lambda}{}_{\mu}
+\partial_{\mu}\xi^{\nu}A^{\Lambda}{}_{\nu}\, ,
\end{equation}

\noindent
and 

\begin{equation}
\label{eq:lambda}
\delta_{\rm  gauge}(\Lambda)A^{\Lambda}{}_{\mu} =
\partial_{\mu}\Lambda^{\Lambda}\, ,
\hspace{1cm}
\Lambda^{\Lambda}\equiv -\xi^{\nu}A^{\Lambda}{}_{\nu}\, .  
\end{equation}

In presence of the functions $f_{\Lambda\Sigma}$, we have 

\begin{equation}
[\delta_{\eta},\delta_{\epsilon}] A_{\Lambda\, \mu} = 
 -2\Re{\rm e}[f^{*}_{\Lambda\Sigma}F^{\Sigma\, -}{}_{\mu\nu}]\xi^{\nu}=
-F_{\Lambda\, \mu\nu}\xi^{\nu} = [\delta_{\rm g.c.t.}(\xi)+\delta_{\rm
  gauge}(\Lambda_{\Lambda})]A_{\Lambda\, \mu}\, ,
\end{equation}

\noindent
where the g.c.t.s and gauge transformations have the same form and the
parameter of the gauge transformations is now

\begin{equation}
\label{eq:lambdadual}
\Lambda_{\Lambda}\equiv \xi^{\nu}A_{\Lambda\, \nu}\, .    
\end{equation}


\subsection{2-forms}
\label{sec-2-forms}

2-forms can be introduced in the theory by dualizing the Noether currents
associated to those isometries of the scalar manifold that are symmetries of
the whole theory \cite{Bergshoeff:2007ij}. We are always talking, then, of a subgroup of
$Sp(2n_{V},\mathbb{R})$ \cite{Gaillard:1981rj}. The action of these
symmetries on the fields is

\begin{eqnarray}
\delta Z^{i} & = & \alpha^{A}k_{A}{}^{i}(Z)\, ,\\
& & \nonumber \\  
\delta \mathcal{F} & = & \alpha^{A}T_{A}\mathcal{F}\, ,\\
\end{eqnarray}

\noindent
where $\mathcal{F}$ is defined in Eq.~(\ref{eq:dualF}) and where the $T_{A}$
are matrices of $\mathfrak{sp}(2n_{V})$ that generate the Lie algebra of the
symmetry group:

\begin{equation}
[k_{A},k_{B}]=-f_{AB}{}^{C}k_{C}\, ,
\hspace{1cm}  
[T_{A},T_{B}]=+f_{AB}{}^{C}T_{C}\, .
\end{equation}

The computation of the Noether current proceeds as in Ref.~\cite{Bergshoeff:2007ij} 
and the result is identical, up to the difference between the period matrix
and $f_{\Lambda\Sigma}$:

\begin{equation}
J_{N\, \mu} = \alpha^{A}J_{N\, A\, \mu}\, ,
\hspace{1cm}
J_{N\, A\, \mu}=
2k^{*}_{A\, i}\partial_{\mu}Z^{i} +\mathrm{c.c.}
-2\langle\, \star \mathcal{F}^{\mu\nu}\mid T_{A} \mathcal{A}_{\nu}\, \rangle \, . 
\end{equation}

These Noether currents are covariantly conserved, i.e.

\begin{equation}
d\star J_{N\, A}=0\, ,  
\end{equation}

\noindent
which implies the local existence of 2-forma $B_{A}$ such that

\begin{equation}
dB_{A}\equiv \star J_{N\, A}= 2k^{*}_{A\, i}\star dZ^{i} +\mathrm{c.c.}
-2\langle\, \mathcal{F}\mid T_{A} \mathcal{A}\, \rangle\, .  
\end{equation}

\noindent
The second term in the r.h.s.~is not invariant under the gauge transformations
of the vector potentials, and the same is therefore true for the 2-forms
$B_{A}$, which transform as

\begin{eqnarray}
\delta_{\rm gauge} \mathcal{A} & = & d\Lambda\, ,\\
& & \nonumber \\
\delta_{\rm gauge}(\Lambda,\Lambda_{1\, A}) B_{A} & = &  d\Lambda_{1\, A}
-2 \langle\, \mathcal{F}\mid T_{A} \Lambda\, \rangle\, .   
\end{eqnarray} 

\noindent
One, then, defines the gauge-invariant 3-form field strengths

\begin{equation}
H_{A}\equiv dB_{A}  +2\langle\, \mathcal{F}\mid T_{A} \mathcal{A}\, \rangle
= 2k^{*}_{A\, i}\star dZ^{i}+\mathrm{c.c.}\, .
\end{equation}

Inspired by the results of Ref.~\cite{Bergshoeff:2007ij} it is not difficult to guess
the form of the supersymmetry transformation rules of these 2-forms:

\begin{equation}
  \begin{array}{rcl}
\delta_{\epsilon}B_{A\, \mu\nu} & = & -{\textstyle\frac{i}{2}}k^{*}_{A\, i}
\bar{\epsilon}_{\bullet}\gamma_{\mu\nu}\chi_{\bullet}{}^{i} +\mathrm{c.c.}\\
& & \\
& & 
+i\mathcal{P}_{A}\bar{\epsilon}^{\bullet}\gamma_{[\mu|}\psi_{\bullet |\nu]} 
+\mathrm{c.c.} \\
& & \\
& & 
-4\langle\, \mathcal{A}_{[\mu|} \mid T_{A}\delta_{\epsilon}\mathcal{A}_{|\nu]}\, \rangle\, ,  \\
\end{array}
\end{equation}

\noindent
where $\mathcal{P}_{A}$ is the momentum map associated to the Killing vector
$k_{A}$.

We find 

\begin{equation}
[\delta_{\eta},\delta_{\epsilon}] B_{A\, \mu\nu } = 
[\delta_{\rm g.c.t.}(\xi)+\delta_{\rm
  gauge}(\Lambda,\Lambda_{1\, A})]B_{A\, \mu\nu }\, ,
\end{equation}

\noindent
where $\xi$ is defined in Eq.~(\ref{eq:xi}), $\Lambda$ in
Eqs.~(\ref{eq:lambda}) and (\ref{eq:lambdadual}) and $\Lambda_{1\, A}$ is
given by

\begin{equation}
\label{eq:lambda1}
\Lambda_{1\, A\, \mu} \equiv -2 \mathcal{P}_{A}\xi_{\mu}\, .  
\end{equation}

A shown in Ref.~\cite{Bergshoeff:2007ij} in $N=2,d=4$ supergravity theories, these
2-forms can be coupled to strings of different species labeled by $A$ whose
tensions would be proportional to $\mathcal{P}_{A}$.


\section{Conclusions}
\label{sec-conclusions}

We have found all the supersymmetric configurations and solutions of ungauged
$N=1,d=4$ with arbitrary couplings to vector and chiral supermultiplets. It is
clear that, qualitatively, these are those of ungauged $N=2,d=4$ supergravity
whose fields and Killing spinors survive the $N=2\, \rightarrow N=1$
truncation explained in the appendix, although the scalar manifolds of the
$N=1$ theory are more general. In particular, all the $N=2$ supersymmetric
configurations in the timelike class (typically black holes) do not survive to
this truncation since their supersymmetry projectors

\begin{equation}
\epsilon_{I}+i\epsilon_{IJ}\gamma_{0}\epsilon^{J}=0\, ,  
\end{equation}

\noindent
involve necessarily the two supersymmetry parameters and one of them is
eliminated in the truncation. The fields of extreme, supersymmetric $N=2,d=4$
black holes may still survive the truncation to $N=1$, but they will not be
BPS in this theory.  

The Killing spinors supersymmetric configurations of the
null class obey projections of the form

\begin{equation}
\gamma^{u}\epsilon_{I}=0\, ,\,\,\,\,\, I=1,2\, ,  
\end{equation}

\noindent
and, thus, they always survive the truncation.

It is likely that the situation in the most general (gauged) $N=1,d=4$ theory
is the same, and, again qualitatively, the supersymmetric solutions can be
obtained by truncation from the $N=2,d=4$ theory on which some partial results
are already available \cite{Huebscher:2007hj}. Of course, a direct calculation
is necessary and, anyway, the most general supersymmetric solutions of gauged
$N=2,d=4$ supergravity are not known, although progress in this direction is
being made \cite{kn:penna}. Work in this direction is already in progress
\cite{kn:HMOV}.

Further extensions (3- and 4-forms) are clearly possible and a more general
study of the possibilities in more general (gauged) $N=1,d=4$ supergravities
has to be performed \cite{kn:BHHO} to compare the results with those of the
Kac-Moody approach.


\section*{Acknowledgments}

This work has been supported in part by the Spanish Ministry of Science and
Education grants FPA2006-00783 and PR2007-0073, the Comunidad de Madrid grant
HEPHACOS P-ESP-00346, the Spanish Consolider-Ingenio 2010 Program CPAN
CSD2007-00042, and by the EU Research Training Network \textit{Constituents,
  Fundamental Forces and Symmetries of the Universe} MRTN-CT-2004-005104. The
author would like to thank the the Stanford Institute for Theoretical Physics
for its hospitality, E.~Bergshoeff, J.~Hartong, M.~H\"ubscher and P.~Meessen
for useful discussions and and M.M.~Fern\'andez for her continuous support.

\appendix


\section{Truncating $N=2$ to $N=1$ supergravity in $d=4$}
\label{app-truncation}

The purpose of this appendix is to show, following
Refs.~\cite{Andrianopoli:2001zh,Andrianopoli:2001gm}, how ungauged $N=2,d=4$ supergravity coupled
to vector multiplets can be truncated to ungauged $N=1,d=4$ supergravity by
decoupling the $N=1$ supermultiplet that contains the second gravitino
$\psi_{2\mu}$. We will only deal with the leading terms in fermions. In doing
so, we will obtain $N=1,d=4$ supergravity in suitable conventions and the
relations between the fields of both theories.


\subsection{Ungauged matter-coupled $N=2,d=4$ supergravity}

We start by a very brief description of ungauged $N=2,d=4$ supergravity
coupled to vector multiplets referring the reader to
Refs.~\cite{Meessen:2006tu,Bellorin:2005zc} for detailed description of the
conventions and further references to the literature.

The gravity multiplet of the $N=2,d=4$ theory consists of the graviton
$e^{a}{}_{\mu}$, a pair of gravitinos $\psi_{I\, \mu}\, ,\,\,\, (I=1,2)$ which
we describe as Weyl spinors, and a vector field $A_{\mu}$.  Each of the $n$
vector supermultiplets of $N=2,d=4$ supergravity that we are going to couple
to the pure supergravity theory contains complex scalar $Z^{i}\, ,\,\,\,\,
(i=1,\cdots, n_{V})$, a pair of gauginos $\lambda^{I\, i}$, which we also
describe as Weyl spinors and a vector field $A^{i}{}_{\mu}$. In the coupled
theory, the $n_{V}+1$ vectors can be treated on the same footing and they
are described collectively by an array $A^{\Lambda}{}_{\mu}\,\,\,\,\,
(\Lambda=1,\cdots,n_{V}+1)$. The coupling between the complex scalars is
described by a non-linear $\sigma$-model with K\"ahler metric
$\mathcal{G}_{ij^{*}}(Z,Z^{*})$ , and the coupling to the vector fields by a
complex scalar-field-valued matrix $\mathcal{N}_{\Lambda\Sigma}(Z,Z^{*})$.
These two couplings are related by a structure called special K\"ahler
geometry, described in the references.Each hypermultiplet consists of 4 real
scalars $q$ (\textit{hyperscalars}) and 2 Weyl spinors $\zeta$ called
\textit{hyperinos}. The $4m$ hyperscalars are collectively denoted by $q^{u}\,
,\,\,\,u=1,\cdots,4m$ and the $2n_{H}$ hyperinos are collectively denoted by
$\zeta_{\alpha}\, ,\,\,\,\alpha=1,\cdots,2n_{H}$.  The $4n_{H}$ hyperscalars
parametrize a quaternionic K\"ahler manifold with metric $\mathsf{H}_{uv}(q)$.

The action for the bosonic fields of the theory is

\begin{equation}
\label{eq:actionN2}
\begin{array}{rcl}
 S & = & {\displaystyle\int} d^{4}x \sqrt{|g|}
\left[R +2\mathcal{G}_{ij^{*}}\partial_{\mu}Z^{i}
\partial^{\mu}Z^{*\, j^{*}} +2\mathsf{H}_{uv}\partial_{\mu}q^{u} \partial^{\mu}q^{v}
 \right. \\
& & \\
& & \left. 
\hspace{2cm}
+2\Im{\rm m}\mathcal{N}_{\Lambda\Sigma} 
F^{\Lambda\, \mu\nu}F^{\Sigma}{}_{\mu\nu}
-2\Re{\rm e}\mathcal{N}_{\Lambda\Sigma} 
F^{\Lambda\, \mu\nu}{}^{\star}F^{\Sigma}{}_{\mu\nu}
\right]\, ,
\end{array}
\end{equation}

\noindent
In these conventions $\Im{\rm m}\mathcal{N}_{\Lambda\Sigma}$ is negative
definite.

For vanishing fermions, the supersymmetry transformation rules of the
fermions are

\begin{eqnarray}
\delta_{\epsilon}\psi_{I\, \mu} & = & 
\mathfrak{D}_{\mu}\epsilon_{I} 
+\epsilon_{IJ}T^{+}{}_{\mu\nu}\gamma^{\nu}\epsilon^{J}\, ,
\label{eq:gravisusyrule}\\
& & \nonumber \\
\delta_{\epsilon}\lambda^{Ii} & = & 
i\not\!\partial Z^{i}\epsilon^{I} +\epsilon^{IJ}\not\!G^{i\, +}\epsilon_{J}\, ,
\label{eq:gaugsusyrule} \\
& & \nonumber \\
\delta_{\epsilon}\zeta_{\alpha} & = &
-i\mathbb{C}_{\alpha\beta}\ \mathsf{U}^{\beta I}{}_{u}\ \varepsilon_{IJ}\ 
\not\!\partial q^{u}\ \epsilon^{J}\, , 
\label{eq:hypersusyrule}
\end{eqnarray}

\noindent
where $\mathfrak{D}_{\mu}$, the Lorentz- and K\"ahler- and $SU(2)$-covariant
derivative acts on the spinors $\epsilon_{I}$ as

\begin{equation}
\mathfrak{D}_{\mu} \epsilon_{I} = 
(\nabla_{\mu} \ +\ {\textstyle\frac{i}{2}}\ \mathcal{Q}_{\mu})\ \epsilon_{I} 
\ +\ \mathsf{A}_{\mu\, I}{}^{J}\ \epsilon_{J}\, .
\end{equation}

\noindent
and $\mathcal{Q}_{\mu}$ is the pullback of the K\"ahler 1-form defined
in Eq.~(\ref{eq:K1form})

\noindent
and $\mathsf{A}_{\mu\, I}{}^{J}$ is the pullback of the $SU(2)$ connection
$\mathsf{A}_{I}{}^{J}$.

\noindent
The 2-forms $T$ and $G^{i}$ are the
combinations

\begin{eqnarray}
T_{\mu\nu} & \equiv & \mathcal{T}_{\Lambda}F^{\Lambda}{}_{\mu\nu}\, ,\\
& & \nonumber \\
G^{i}{}_{\mu\nu} & \equiv &
\mathcal{T}^{i}{}_{\Lambda}F^{\Lambda}{}_{\mu\nu}\, ,
\end{eqnarray}

\noindent
where, in turn, $\mathcal{T}_{\Lambda}$ and $\mathcal{T}^{i}{}_{\Lambda}$ are,
respectively, the graviphoton and the matter vector fields projectors, defined
by

\begin{eqnarray}
\label{eq:projectorg}
\mathcal{T}_{\Lambda} & \equiv & 2i \mathcal{L}_{\Lambda}
=2i\mathcal{L}^{\Sigma}\Im{\rm m}\, 
\mathcal{N}_{\Sigma\Lambda}\, ,\\
& & \nonumber \\
\label{eq:projectorm}
\mathcal{T}^{i}{}_{\Lambda} & \equiv & -f^{*}{}_{\Lambda}{}^{i}=
-\mathcal{G}^{ij^{*}}
f^{*\, \Sigma}{}_{j^{*}}\Im{\rm m}\, 
\mathcal{N}_{\Sigma\Lambda}\, .
\end{eqnarray}

The supersymmetry transformations of the bosons are

\begin{eqnarray}
  \delta_{\epsilon} e^{a}{}_{\mu} & = & 
-{\textstyle\frac{i}{4}} (\bar{\psi}_{I\, \mu}\gamma^{a}\epsilon^{I}
+\bar{\psi}^{I}{}_{\mu}\gamma^{a}\epsilon_{I})\, ,
\label{eq:susytranse}\\
& & \nonumber \\ 
  \delta_{\epsilon} A^{\Lambda}{}_{\mu} & = & 
{\textstyle\frac{1}{4}}
(\mathcal{L}^{\Lambda\, *}
\epsilon^{IJ}\bar{\psi}_{I\, \mu}\epsilon_{J}
+
\mathcal{L}^{\Lambda}
\epsilon_{IJ}\bar{\psi}^{I}{}_{\mu}\epsilon^{J}) \nonumber \\
& & \nonumber \\
& & 
+
{\textstyle\frac{i}{8}}(f^{\Lambda}{}_{i}\epsilon_{IJ}
\bar{\lambda}^{Ii}\gamma_{\mu}
\epsilon^{J}
+
f^{\Lambda *}{}_{i^{*}}\epsilon^{IJ}
\bar{\lambda}_{I}{}^{i^{*}}\gamma_{\mu}\epsilon_{J})\, ,
\label{eq:susytransA}\\
& & \nonumber \\
  \delta_{\epsilon} Z^{i} & = & 
{\textstyle\frac{1}{4}} \bar{\lambda}^{Ii}\epsilon_{I}\, ,
\label{eq:susytransZ}\\
& & \nonumber \\
  \delta_{\epsilon} q^{u} & = & 
{\textstyle\frac{1}{4}}\mathsf{U}_{\alpha I}{}^{u} 
(\bar{\zeta}^{\alpha}\epsilon^{I} 
+\mathbb{C}^{\alpha\beta}\epsilon^{IJ}\bar{\zeta}_{\beta}\epsilon_{J})\, .
\label{eq:susytransq}
\end{eqnarray}


\subsection{Truncation to $N=1,d=4$ supergravity}

The truncation to $N=1,d=4$ supergravity consists in setting to zero the
supermultiplet that contains the second gravitino $\psi_{2\, \mu}$ and the
graviphoton. The remaining fields in the supergravity multiplet
$\{e^{a}{}_{\mu},\psi_{1\mu}\}$ will become those of the $N=1,d=4$
supergravity multiplet and the $n_{V}$ $N=2,d=4$ vector multiplets will be
split into $n_{V}$ chiral multiplets, each of them containing one complex
scalar and the first component of one $N=2$ gaugino $\lambda^{1 i}$ and
$n_{V}$ vector multiplets, each of them containing one vector and the second
component of one $N=2$ gaugino $\lambda^{2 i}$. However, not all of them can
simultaneously. Finally, only half of the hyperscalars, parametrizing a
K\"ahler manifold will survive the truncation.

We relabel the $N=2$ indices $\Lambda \rightarrow \mathbf{\Lambda}$ and
$i,i^{*}\rightarrow \mathbf{i},\mathbf{i^{*}}$ to label the $N=1$ vector
multiplets with $\Lambda$ and the chiral multiplets with $i$. We set 

\begin{equation}
\psi_{2\, \mu}=\delta_{\epsilon} \psi_{2\, \mu}=\epsilon_{2}=0\, , 
\end{equation}

\noindent
and define

\begin{equation}
\psi_{\bullet\, \mu}\equiv \psi_{1\, \mu}\, ,
\hspace{1cm}
\epsilon_{\bullet}\equiv \epsilon_{1}\, .
\end{equation}

\noindent
The supersymmetry transformations of the two gravitini become

\begin{eqnarray}
\delta_{\epsilon}\psi_{\bullet\, \mu} & = & 
\left(\nabla_{\mu} +{\textstyle\frac{i}{2}} \mathcal{Q}_{\mu} 
+\mathsf{A}_{\mu\, 1}{}^{1}\right)\epsilon_{\bullet} \, ,
\label{eq:gravisusyrule1}\\
& & \nonumber \\
\delta_{\epsilon}\psi_{2\, \mu} & = & \mathsf{A}_{\mu\, 1}{}^{2}\epsilon_{\bullet}
-T^{+}{}_{\mu\nu}\gamma^{\nu}\epsilon^{\bullet}=0\, .
\label{eq:gravisusyrule2}
\end{eqnarray}

This means that the component $\mathsf{A}_{\mu\, 1}{}^{1}$ of the $SU(2)$
connection has to be integrated into the K\"ahler connection and the component
$\mathsf{A}_{\mu\, 1}{}^{2}$ and the graviphoton field strength has to be set
to zero

\begin{eqnarray}
\mathsf{A}_{\mu\, 1}{}^{2} &  = & 0\, ,\\
& & \nonumber \\
\label{eq:Tconstraint}
T^{+}{}_{\mu\nu} & = & 0\, .  
\end{eqnarray}

The supersymmetry transformation rule of the graviton becomes, simply

\begin{equation}
  \delta_{\epsilon} e^{a}{}_{\mu} = 
-{\textstyle\frac{i}{4}} \bar{\psi}_{\bullet\, \mu}\gamma^{a}\epsilon^{\bullet}
+\mathrm{c.c.}
\end{equation}

Let us now consider the $N=2$ vector multiplets.  

The most general solution to the constraint Eq.~(\ref{eq:Tconstraint}) is to
see it as an orthogonality condition between the graviphoton projector and the
vector fields \cite{Andrianopoli:2001zh,Andrianopoli:2001gm}. The $N=2$ vector
index is split $\mathbf{\Lambda}=(\Lambda,X)$, where $\Lambda=1,\cdots,n$ and
$X=0,1,\cdots,\mathbf{n_{V}}-n_{V}\equiv n_{C}$ and

\begin{equation}
\mathcal{T}_{\Lambda}=
2i\mathcal{L}^{\mathbf{\Sigma}}\Im{\rm m}\, 
\mathcal{N}_{\mathbf{\Sigma}\Lambda}
=0\, ,
\hspace{1cm}
F^{X+}{}_{\mu\nu}=0\, .  
\end{equation}

The $N=2$ vector multiplets in the range $\Lambda$ give only $N=1$ vector
multiplets (the chiral multiplets have to be truncated) and those in the range
$X$ give only chiral $N=1$ multiplets (the $N=1$ vector multiplets must be
truncated). Since the dual vector field strengths 

\begin{equation}
F_{X}{}^{+}{}_{\mu\nu}
=\mathcal{N}_{X\mathbf{\Lambda}}F^{\mathbf{\Lambda}+}{}_{\mu\nu} 
=\mathcal{N}_{X\Lambda}F^{\Lambda+}{}_{\mu\nu} 
+\mathcal{N}_{XY}F^{Y+}{}_{\mu\nu}\, , 
\end{equation}

\noindent
must also vanish for consistency, the off-diagonal blocks of the period matrix
must also vanish

\begin{equation}
\mathcal{N}_{X\Lambda}=0\, ,  
\end{equation}

\noindent
and, therefore

\begin{equation}
\mathcal{T}_{\Lambda}=
2i\mathcal{L}^{\Sigma}\Im{\rm m}\, 
\mathcal{N}_{\Sigma\Lambda}
=0\,\,\, \Rightarrow \mathcal{L}^{\Lambda}=0\, .  
\end{equation}

\noindent
Only the components $\mathcal{L}^{X}$ survive, and, together with the period
matrix $\mathcal{N}_{XY}$, define a special K\"ahler manifold of dimension
$\mathbf{n_{V}}-n_{V}=n_{C}$ and with K\"ahler metric

\begin{equation}
\mathcal{G}_{ij^{*}} = 
-2\Im{\rm m}\,
\mathcal{N}_{XY}f^{X}{}_{i}f^{Y}{}_{j^{*}}\, ,
\hspace{1cm}
i,j=1,\cdots,n_{C}\, .  
\end{equation}

\noindent
The diagonal block 

\begin{equation}
\mathcal{N}_{\Lambda\Sigma}\equiv
{\textstyle\frac{1}{2}}f^{*}_{\Lambda\Sigma}\, ,
\end{equation}

\noindent
determines the couplings of the scalars of the chiral multiplets to the
vectors. It can be shown that $f_{\Lambda\Sigma}$ is a holomorphic function of
the $Z^{i}$s. 

The consistency of these conditions leads to several conditions that the
special K\"ahler manifold has to satisfy on order to be reducible to $N=1$ and
can be found in \cite{Andrianopoli:2001zh,Andrianopoli:2001gm}.

It is convenient to study the supersymmetry transformations of the two gaugini
in the form

\begin{equation}
f^{\mathbf{\Lambda}}{}_{\mathbf{i}}\delta_{\epsilon}\lambda^{I\mathbf{i}}
=if^{\mathbf{\Lambda}}{}_{\mathbf{i}}\not\!\partial Z^{\mathbf{i}}\epsilon^{I}
+{\textstyle\frac{1}{2}}\not\!
F^{\mathbf{\Lambda}+}\epsilon^{IJ}\epsilon_{J}\, ,
\end{equation}

\noindent
where we have used the constraint Eq.~(\ref{eq:Tconstraint}).  Then, splitting
the index $\mathbf{i}=(\alpha,i)$ with $\alpha=1,\cdots,n$ and $i=1,\cdots,
n_{C}$, the above equation splits as follows

\begin{eqnarray}
f^{\Lambda}{}_{\mathbf{i}}\delta_{\epsilon}\lambda^{1\mathbf{i}} & = & 
0\, ,
\label{eq:gaugsusyrull1} \\
& & \nonumber \\
f^{X}{}_{i}\delta_{\epsilon}\lambda^{1i} & = & 
if^{X}{}_{i}\not\!\partial Z^{i}\epsilon^{\bullet}\, ,
\label{eq:gaugsusyruli1} \\
& & \nonumber \\
f^{\Lambda}{}_{\alpha}\delta_{\epsilon}\lambda^{2\alpha} & = & 
{\textstyle\frac{1}{2}}\not\!
F^{\mathbf{\Lambda}+}\epsilon_{\bullet}\, ,
\label{eq:gaugsusyrulel2}\\ 
& & \nonumber \\
f^{X}{}_{i}\delta_{\epsilon}\lambda^{2i} & = & 
0\, .
\label{eq:gaugsusyrule2} 
\end{eqnarray}

\noindent
Then, we define the $N=1$ gaugini and dilatini

\begin{eqnarray}
\lambda_{\bullet}{}^{\Lambda} & \equiv & 
-f^{\Lambda}{}_{\alpha}\lambda^{2\alpha}\, ,  \\
& & \nonumber \\
\chi_{\bullet}{}^{i} & \equiv & \lambda^{1i}\, ,  
\end{eqnarray}

\noindent
and set to zero all the other components. Their resulting supersymmetry
transformation rules are 

\begin{eqnarray}
\delta_{\epsilon}\lambda_{\bullet}{}^{\Lambda} & = &
{\textstyle\frac{1}{2}}\not\!
F^{\mathbf{\Lambda}+}\epsilon_{\bullet}\, ,\\
& & \nonumber \\
\delta_{\epsilon}\chi_{\bullet}{}^{i} & = & 
i\not\!\partial Z^{i}\epsilon^{\bullet}\, .
\end{eqnarray}

The supersymmetry transformation rules of the vector fields are split in 

\begin{eqnarray}
\label{eq:susytransAl}
  \delta_{\epsilon} A^{\Lambda}{}_{\mu} & = & 
{\textstyle\frac{i}{8}}
\bar{\lambda}_{\bullet}{}^{\Lambda}\gamma_{\mu}
\epsilon^{\bullet}
+\mathrm{c.c.}\, ,\\
& & \nonumber \\
\label{eq:susytransAX}
  \delta_{\epsilon} A^{X}{}_{\mu} & = & 0\, .
\end{eqnarray}

Finally, the supersymmetry transformation rules of the scalars split into

\begin{eqnarray}
  \delta_{\epsilon} Z^{i} & = & 
{\textstyle\frac{1}{4}} \bar{\chi}_{\bullet}{}^{i}\epsilon_{\bullet}\, ,
\label{eq:susytransZi}\\
& & \nonumber \\
  \delta_{\epsilon} Z^{\alpha} & = & 0\, .
\label{eq:susytransZa}
\end{eqnarray}

Let us now consider the truncation in the hypermultiplet sector.  The $4n_{H}$
real dimensional quaternionic-K\"ahler manifold has to be truncated to a
$n_{H}$ complex dimensional K\"ahler manifold
\cite{Andrianopoli:2001zh,Andrianopoli:2001gm}. The truncation can only be
done in some quaternionic-K\"ahler manifold: if we split the $Sp(2n_{H})$
index $\alpha$ into $A,\dot{A}=1,\cdots ,n_{H}$ and the undotted indices
correspond to the sector which will survive, the components
$\Omega_{\dot{A}\dot{B}\dot{C}\dot{D}}$ must vanish identically. If this
condition is satisfied, then one can set 

\begin{equation}
\mathsf{U}^{2A} = \mathsf{A}^{1}  =  \mathsf{A}^{2} =
\Delta^{A}{}_{\dot{B}}=\zeta_{\dot{A}}=0\, ,
\end{equation}

\noindent
consistently. The surviving components of the Quadbein are $\mathsf{U}^{1A}$
and its complex conjugate $\mathsf{U}^{2\dot{A}}$ which can be expressed in
terms of just $n_{H}$ holomorphic coordinates $w^{s}$. 

The independent non-vanishing supersymmetry transformation rules of the
hyperscalars and the hyperinos are

\begin{eqnarray}
(\mathsf{U}_{1Au})^{*}\delta_{\epsilon} q^{u} & = & 
{\textstyle\frac{1}{4}}\bar{\zeta}^{\bullet\, A}\epsilon^{\bullet}\, ,\\
& & \nonumber \\
\mathsf{U}^{1Au}\delta_{\epsilon} \zeta_{\bullet A} & = & 
i\not\!\partial q^{u} \epsilon^{\bullet}\, .
\end{eqnarray}

\noindent
Using the holomorphic coordinates $w^{s}$ we now define the $n_{H}$ $N=1$
dilatini $\zeta^{s}$

\begin{equation}
\zeta_{\bullet}{}^{s}\equiv\mathsf{U}^{1As} \zeta_{\bullet\, A}\, ,  
\end{equation}

\noindent
and the above supersymmetry transformation rules take the standard form

\begin{eqnarray}
\delta_{\epsilon} w^{s} & = & 
{\textstyle\frac{1}{4}}\bar{\zeta}_{\bullet}{}^{s}\epsilon_{\bullet}\, ,\\
& & \nonumber \\
\delta_{\epsilon} \zeta_{\bullet}{}^{s} & = & 
i\not\!\partial w^{s} \epsilon^{\bullet}\, .
\end{eqnarray}

The quaternionic K\"ahler manifolds that can be truncated to $N=1$ chiral
multiplets are precisely those in which one can construct cosmic string
solutions (\textit{hyperstrings}): in Ref.~\cite{Huebscher:2006mr} the
supersymmetry equations were solved by choosing a metric of  the form

\begin{equation}
ds^{2}= dt^{2}-(dx^{3})^{2}
-2e^{\Phi(z,z^{*})}dzdz^{*}\, ,
\end{equation}

\noindent
hyperscalars which are real functions of the complex coordinate $z$ and its
complex conjugate $q^{u}(z,z^{*})$. In these conditions, the supersymmetry
conditions take the form

\begin{eqnarray}
\label{eq:embeddingagain1}
 \mathsf{U}^{\alpha 2}{}_{u}
\partial_{\underline{z}}q^{u}
=
\mathsf{U}^{\alpha 1}{}_{u}
\partial_{\underline{z}^*}q^{u}
&  = &  0\, , \\
& & \nonumber \\
\label{eq:embeddingagain2}
\varpi_{\underline{z}}{}^{zz^{*}} & = &
\mathsf{A}^{3}{}_{u}\ \partial_{\underline{z}}q^{u}\, ,\\
& & \nonumber \\
\label{eq:embeddingagain3}
\mathsf{A}^{1}{}_{u}\ \partial_{\underline{m}}q^{u} =
\mathsf{A}^{2}{}_{u}\ \partial_{\underline{m}}q^{u} & = & 0\, .
\end{eqnarray}

They are clearly solved by setting $\mathsf{A}^{1}{}_{u}=\mathsf{A}^{2}{}_{u}=
\mathsf{U}^{\alpha 2}{}_{u}=0$ and, then, taking the hyperscalars to be
holomorphic functions of $z$ $\partial_{\underline{z}^*}q^{u}=0$.


\end{document}